\documentstyle[twocolumn,aps,prl,epsf,floats]{revtex}
 \def\half{\mbox{$1\over 2$}}

 \def\beq{\begin{equation}}
 \def\eeq{\end{equation}}
 \def\beqa{\begin{eqnarray}}
 \def\eeqa{\end{eqnarray}}
 
 \def\LP{\left(}
 \def\RP{\vphantom{\half} \right)}

 \gdef\aver#1{\left\langle #1 \right\rangle}
 \gdef\s#1{\! #1 \!}
 \gdef\l#1{\> #1 \>}
 \gdef\Eq#1{Eq.~(\ref{#1})}

\begin{document}
\draft

\preprint{NSF-ITP-97-070}

\title{
Non-equilibrium Josephson--like effects
\ifpreprintsty \\ \fi 
in mesoscopic S-N-S junctions}

\author{Nathan Argaman}
\address{ Institute for Theoretical Physics,
University of California, Santa Barbara, CA 93106, USA }

\address{\bigskip\rm
(August 29, 1997)}

\address{\parbox{14.5cm}{\rm\small
\bigskip
Wide mesoscopic superconducting--normal-metal--superconducting (S-N-S)
junctions exhibit Andreev bound states which carry substantial
supercurrents, even at temperatures for which the equilibrium
Josephson effect is exponentially small --- the currents carried by
different states can cancel each other.  This cancellation is
incomplete whenever the junctions are driven out of equilibrium, e.g.,
by a dc voltage.  This leads to phenomena similar to the usual dc
and ac Josephson effects, but dominated by the second harmonic of the
Josephson frequency, which may explain some striking recent
experiments.  A simple description of these, in the spirit of the
Resistively--Shunted--Junction model, is suggested.
}}

\address{\parbox{14.5cm}{\bigskip \rm \small
}
}
\maketitle

\begin{figure}[tb]
\epsfxsize=0.9\hsize
\centerline{ \epsffile{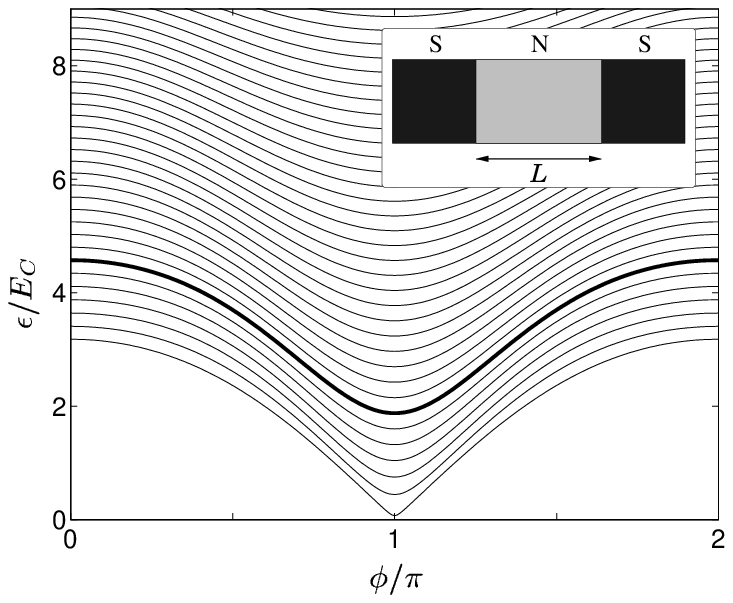} \quad}
\refstepcounter{figure}
\label{fig_dos}
FIG. \ref{fig_dos}: 
Energies of some of the Andreev bound states $E_n$ (equally spaced in
$n$), as a function of $\phi$, for a diffusive S-N-S junction (see
inset) with $\delta \ll E_C \ll \Delta$ (and $\tau_\varphi \to \infty$).  
Thick line: a representative used below, $E_{\rm rep}(\phi)$.
\end{figure}

Mesoscopic superconducting--normal-metal--supercon\-ducting (S-N-S)
junctions exhibit a spectrum of low--lying electronic ``Andreev bound
states'', which depends strongly on the phase--difference $\phi$
between the order parameters in S (see Fig.~\ref{fig_dos}).
This $\phi$--dependence persists when the temperature $T$ is raised,
and the ``normal--metal coherence length'', 
$\xi_N = \sqrt{\hbar D / k_B T}$, becomes much smaller than the
distance $L$ between the two S electrodes ($D$ is the diffusion
constant in N).  In this high--temperature regime the equilibrium
Josephson coupling is exponentially weak.  However, as the
spectrum depends on $\phi$, and hence on time $t$ (the voltage $V$ is 
proportional to $d\phi/dt$ ), non--equilibrium (NEQ) effects occur,
including supercurrents which are the topic of this Letter.

Such junctions have attracted much attention.  The $\phi$--dependence
of their conductance was studied extensively in recent years
\cite{Courtois}, and it was demonstrated that a naive Ginzburg--Landau
description of proximity effects was insufficient.  In these ``Andreev
interferometers'' N is connected to external normal--metal electrodes,
which thermalize the occupations \cite{Volkov}.  In contrast, for
an ``isolated'' N the electronic excitations are ``bound'': an 
electron near the Fermi level in N cannot enter S because of the
superconducting gap, except by ``Andreev reflection'' --- it pairs
with another electron from the Fermi sea, leaving behind a hole.
Conversely, a hole may break up a Cooper pair and produce an electron.
These processes coherently mix the electron and hole states in N, in a
$\phi$--dependent manner.  As energy--relaxation is slow in the
mesoscopic regime ($\tau_\varphi > L^2/D$, with $\tau_\varphi$ the 
single--particle dephasing time), NEQ situations naturally develop.

NEQ phenomena have been studied for both clean
\cite{Wurzburg,Shumeiko} and dirty \cite{Averin,Spivak} S-N-S
junctions \cite{Tinkham}.  The progress made here is in (a) developing a
simple but versatile model of NEQ phenomena, in the spirit of the
resistively--shunted--junction (RSJ) model (at the price of a
restriction to small voltages); and (b) using the unusual harmonic
content of the NEQ supercurrents to understand a recent observation
\cite{Allen}, which cannot be explained \cite{comment} in terms of
previously proposed effects \cite{SpiKhmel}.
The simplest description of NEQ, with a constant energy--relaxation
rate $1/\tau_E$ (with $\tau_E \geq \tau_\varphi$), is used to discuss
wide (or long) dirty junctions, in the high--temperature regime, 
$E_C = \hbar D/L^2 \ll T \ll \Delta$.
Here $E_C$ is the Thouless energy (the correlation energy of the
spectrum), $k_B=1$, and $\Delta$ is the gap in S.  The
single--particle level spacing in N, $\delta$, is taken small
(metallic limit): $\delta \ll \hbar/\tau_E \ll E_C$.

The density of states $\nu(\epsilon,\phi)$ can be calculated from the
(disorder--averaged) Usadel equations \cite{Argaman,ZS}.  It is
convenient to define the energy $E_n(\phi)$ of the $n$th level: 
$n = \int_0^{E_n} \nu(\epsilon,\phi) \; d\epsilon$,
see Fig.~\ref{fig_dos} ($\epsilon=0$ is the Fermi level).
We have assumed for simplicity that time--reversal symmetry is
preserved, that the pair potential $\Delta$ is constant in S and
vanishes in N, and that the N-S boundaries are ``perfect'', i.e.\ that
no normal reflection occurs and the amplitude for Andreev reflections
is of unit magnitude.  A mini--gap is proximity--induced in N.  Its
size is $E_g \simeq 3.1 E_C$ for $\phi=0$, and it closes when
$\phi=\pi$ and reopens periodically \cite{ZS} (fluctuations in $E_g$
and $E_n$ are only $\sim \delta$).  The spectrum is very sensitive to
symmetry breaking --- for example, a moderate amount of spin--flip
scattering, with $\hbar/\tau_{sf} \sim E_C$, can close the
mini--gap.  However, for most S-N-S geometries it contains a band of
width $\sim E_C$ of levels with a significant $\phi$--dependence.
For our system, the curve 
$E_{\rm rep} \simeq 3.4 E_C \sqrt{1 \s+ 0.7 \cos(\phi)}$ represents
the low--lying part of the spectrum well, and the effective number of 
levels in this band is $N \simeq 10 E_C/\delta$, including a
factor of 2 for spin [a simple shift in energy does not affect the
contribution of a level $E_n(\phi)$].  As we will use only the fact
that $E_{\rm rep}$ is $\phi$--periodic, the results will be
qualitatively applicable also to the experimentally--relevant clean
case, with a mean--free--path of a few times $L$ (though perhaps not
to the theoretical clean limit \cite{Wurzburg,Shumeiko}, which has a 
separable spectrum).

The states $E_n$ carry currents, $I = (2e/\hbar) (dE_n/d\phi)$ per
occupied state, as seen by equating the spent and stored energies, 
$IV dt = \sum dE_n$ (neglecting any changes in the interaction
energy), and using the Josephson relationship $d\phi = (2e/\hbar) V dt$.  
For each $n$ there is also a state at $-E_n$, with the opposite
current.  The total supercurrent is thus
\beq \label{Isum}
I_S = - {2e \over \hbar} \sum_n {dE_n \over d\phi} (1 \s- 2f)  \; ,
\eeq
where $f$ is the occupation probability of $E_n$, and $1 \s- f$ is
that of $-E_n$ (Ref.~\cite{imbal}; the spin index is included in $n$).

The fact that NEQ occupations often enhance supercurrents was
demonstrated in the seventies \cite{Tinkham}, and follows from an
elegant argument.  In thermal equilibrium, 
$f = f_{\rm eq}=1/\LP 1 \s+ \exp(\epsilon/T) \RP$, and $I_S$ takes the
form
\beq \label{Ieq}
I_{\rm eq}(\phi) =  \int_0^\infty d\epsilon \, 
           j(\epsilon,\phi) \, \tanh(\epsilon/2T)  \; ,
\eeq
where $j(\epsilon)$ is the ``Josephson current density'', 
$j(E_n,\phi) \propto \nu(E_n,\phi) (dE_n/d\phi)$.
This $j(\epsilon)$ is the imaginary part of a ``Green's function''
which is analytic in the upper half of the complex $\epsilon$ plane
---  a positive imaginary part of $\epsilon$ corresponds to a
dephasing rate ($\hbar/\tau_\varphi$) which would smooth out any
singularities.  Using contour integration (the integrand is even), one
finds the well--known Matsubara sum:
$I_{\rm eq} = 2 \pi i T \sum_1^\infty j(i\omega_n)$, where 
$\omega_n = (2n \s- 1) \pi T$ are the poles of the $\tanh$ factor.
At high temperatures, even the smallest Matsubara frequency has 
a decay time $\hbar/\omega_1 < L^2/D$, yielding an exponentially small
$j(i\omega_n)$.  Thus, for physical quantities of this form,
thermal averaging is mathematically equivalent to dephasing.

In NEQ situations contour integration cannot be used, and the currents
are not exponentially small.  For \Eq{Ieq} to give small results, 
$j(\epsilon)$ must oscillate: in our diffusive system, $j(i\omega)$
decays exponentially with $\sqrt{\omega/E_C}$ on the 
imaginary axis, and correspondingly $j(\epsilon)$ oscillates and
decays rapidly with $\sqrt{\epsilon/E_C}$.  The $E_n(\phi)$ curves in 
Fig.~\ref{fig_dos} thus change their character repeatedly at higher
$\epsilon$, occasionally having shallow maxima at $\phi = \pi$, rather
than minima.  As can be seen from \Eq{relax} below, the
deviations of $f$ from equilibrium change sign in rhyme with these
oscillations, and thus the integrand of the NEQ part of the supercurrent, 
$2\int_0^\infty d\epsilon \, j \, (f_{\rm eq} \s- f)$,
does not change sign and cannot be affected by any cancellations.

The expression for $I_S$ may thus be approximated by
\beq \label{rep}
I_S  \l\simeq  I_{\rm eq}(\phi) + 2 {2e \over \hbar} N 
     {dE_{\rm rep} \over d\phi} \, (f \s- f_{\rm eq})  \; ,
\eeq
The errors incurred here \cite{check} are probably smaller than those
of the relaxation--time approximation,
\beq \label{relax}
{df \over dt} = -{1 \over \tau_E} (f-f_{\rm eq})  \; ,
\eeq
which we shall also employ.  For example, the latter ignores the
effects of the mini--gap on the electron--electron and
electron--phonon relaxation processes.

\begin{figure}[tb]
\epsfxsize=\hsize 
\epsffile{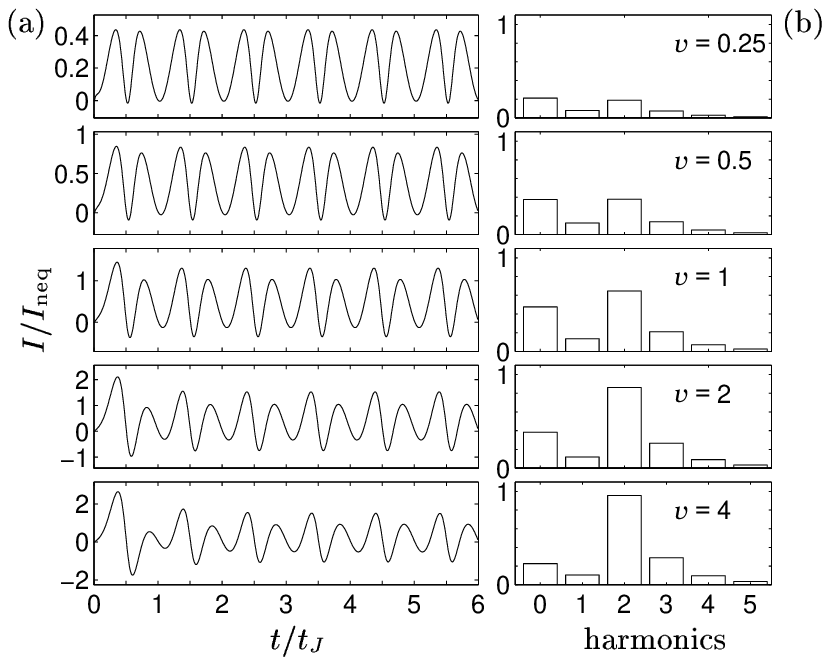}
\vspace*{1.5mm}
\refstepcounter{figure}
\label{fig_ac}
FIG. \ref{fig_ac}:
(a) $I_S$ vs.\ $t$, and (b) amplitudes $I_k$ of its
harmonic decomposition at $t \gg \tau_E$ (with $k \geq 0$;
phases not shown), for five dc voltages labeled by 
$v = (2e/\hbar) V \tau_E$.
\end{figure}

When a dc voltage is applied to the junction,
${d\phi / dt} = 2eV/\hbar = {\rm const.}$, this model gives ac
supercurrents, see Fig.~\ref{fig_ac} (the units $I_{\rm neq}$ and
$t_J$ are defined below).
Here $T \gg E_C$ was used to take $I_{\rm eq}=0$ 
and $f_{\rm eq}=\half-E_{\rm rep}/4T$.  The second harmonic dominates
because $I_S$ of \Eq{rep} is a product of two oscillatory
factors, ${dE_{\rm rep} / d\phi}$ and $(f \s- f_{\rm eq})$.  For junctions
with $\tau_\varphi \sim L^2/D$ one expects a much ``softer''
spectrum, with $E_{\rm rep} \propto \cos(\phi)$, and $I_S$ would
exhibit true frequency doubling, evolving from a $\sin^2(\phi)$
behavior to $-\sin(2\phi)$ as $V$ is increased.

In the limit of small voltages, we reproduce the results of
Refs.~\cite{Spivak,Tinkham}: 
$f \s- f_{\rm eq} \l\simeq - \tau_E (2eV/\hbar)(df_{\rm eq}/d\phi)$,
and
\beq \label{lowV}
I_S =  \left( {2e\over \hbar} \right)^2 {N \tau_E V \over 2T} 
       \left( {dE_{\rm rep} \over d\phi} \right)^2 + O(V^2)  \; .
\eeq
This adds a phase--dependent term to the ohmic conductance of the
junction.  For a large dc voltage, we may 
approximate $f$ by a constant --- the phase average of $f_{\rm eq}$
--- which yields a purely oscillatory $I_S$.   As can be seen from the
``Debye mechanism'' of Ref.~\cite{Spivak}, which monitors the
steady--state transfer of energy into the heat bath ($\propto
1/\tau_E$) rather than that into the electrons, the total power
dissipated approaches a constant $P_{\rm max}$ at large $V$,
and so $\int I \, dt$ decreases as $P_{\rm max}/V$.

So far, we have ignored the normal current component, 
$I_N \simeq G_N V$, where $G_N$ is the conductance of N in the absence
of proximity effects.  Clearly, in the metallic limit the
occupations can not evolve adiabatically (as in Ref.~\cite{Averin}).
Instead $f$ diffuses in energy \cite{Wilkinson}, with a coefficient 
$D_E(\epsilon,\phi) \sim G_N V^2 \delta$.  
To see this, note that in a time $t$ a particle can
cross the junction $\sim t E_C/\hbar$ times, shifting in
energy by $\pm eV$ at each Andreev reflection, so that 
$D_E \sim (eV)^2 D/L^2$ [recall that 
$G_N \sim (e^2/h) E_C/\delta$].
This additional relaxation term can 
be neglected, compared to the terms kept in \Eq{relax}, if 
$\tau_E D_E \ll E_C^2$ or $t_J D_E \ll E_C^2$, depending on
the context ($t_J = h/2eV$ is the Josephson period).  This limits
our model to small voltages, $eV \ll \sqrt{E_C \hbar / \tau_E}$, or 
$eV \ll E_C$, respectively.  It is difficult to avoid the latter
restriction, because at still higher voltages the temporal and spatial
dependencies of the relevant Green's functions become intertwined.

%
\begin{figure}[t]
\epsfxsize=0.9\hsize 
\centerline{ \epsffile{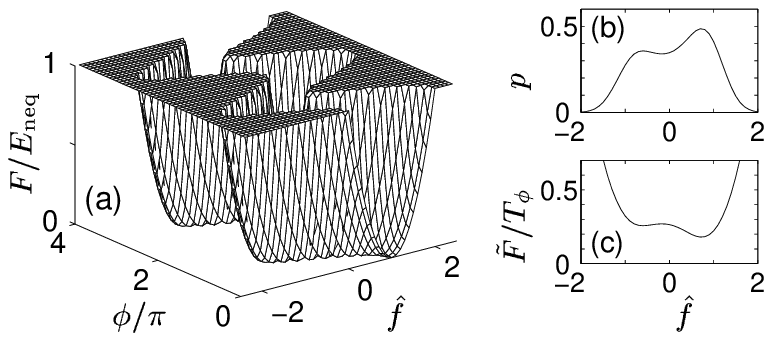} }
\vspace*{1mm}
\refstepcounter{figure}
\label{fig_mesh}
FIG. \ref{fig_mesh}:
(a)  The ``free--energy'' function $F(\phi,\hat f)$, for $I=0$, is
a ``winding valley'' with minimum at $\hat f = \hat E(\phi)$
and a parabolic cross--section.  (b) The probability density 
in $\hat f$, for $\tau_E \to \infty$, is
$p(\hat f) \propto \int d \phi \, \exp(-F/T_\phi)$; here 
$T_\phi = 0.25 E_{\rm neq}$.  (c) The effective potential, 
$\tilde F(\hat f) = -T_\phi \log p$.  When $T \ll T_\phi$, its minima
``trap'' the system.
\end{figure}
In many situations $I(t)$ rather than $V(t)$ is given.  As in the
well--known RSJ model \cite{Likharev}, we write
$I = I_S + I_N + I_F$, where $I_F$ is a fluctuating noise
component (S-N-S junctions are generally overdamped, with no 
displacement current).  We rescale $E_{\rm rep}(\phi)$ and $f$, giving 
\beqa \label{mod1}
{d\phi \over dt}  & \l= &  {2e \over \hbar G_N}
\LP I - I_{\rm neq} 2 { d\hat E \over d\phi } (\hat E-\hat f) -I_F \RP \\
{d\hat f \over dt}  & \l= &  {1 \over \tau_E}
\LP \hat E - \hat f + J_F \RP  \; ,        \label{mod2}
\eeqa
where $\hat E$ and $\hat f$ are defined by 
$E_{\rm rep}(\phi) = A \hat E +B$ and $f = \half - (A \hat f +B)/4T$, 
with the requirement $-1 \leq \hat E \leq 1$ fixing $A$ and $B$ 
(in our model $A\simeq 1.3 E_C$).  Here 
$I_{\rm neq}=(2e/\hbar) N A^2/4T$, and the energy scale is 
$E_{\rm neq} = (\hbar/2e)I_{\rm neq} \simeq 4.1E_C^3/\delta T$,
only a factor of $E_C / 3.5 T$ smaller than the $T=0$
Josephson coupling energy.  The correlators 
$\aver{ I_F(t) I_F(0) } = 2 T_\phi G \, \delta(t)$ and 
$\aver{ J_F(t) J_F(0) } = (\tau_E T / E_{\rm neq}) \, \delta(t)$
give the Gaussian fluctuations of $I_F$ and $J_F$.
For thermal noise, $T_\phi=T$; high--frequency external noise 
can be described by $T_\phi > T$.

Eqs.~(\ref{mod1}) and (\ref{mod2}) describe overdamped motion in a 
free--energy \cite{free_energy} landscape 
$F(\phi,\hat f) =  E_{\rm neq} (\hat E-\hat f)^2 - (\hbar/2e) I \phi$,
see Fig.~\ref{fig_mesh}(a).
This generalizes the tilted washboard potential of the RSJ model,
$F(\phi) = - (\hbar/2e) (I_c \cos \phi+I\phi)$ (where $f$ need not be
followed).  For fixed $\phi$, the fluctuations induced by $J_F$
reproduce the $\sqrt{N}$ noise of thermal occupations (and result in
increased current noise, $I_S \propto \partial F/\partial\phi$,
Ref.~\cite{fluc-diss}).  Note that for $\tau_E \to 0$ (fast
equilibration), $\hat f$ may be integrated out, which must
reproduce the RSJ model.  Here $I_c=0$, but the $\phi$--dependence
of the ``bottom of the valley'' re-emerges at lower $T$.

In the opposite limit, $\tau_E \to \infty$, the ``fast variable''
$\phi$ may be removed (see Fig.~\ref{fig_mesh}), giving a two--peaked 
probability density in $\hat f$ (one broad peak for large
$T_\phi$).  The corresponding effective potential, 
$\tilde F(\hat f)$, has valleys with a depth roughly $\sim T_\phi$,
which trap the system if $T<<T_\phi$ (Ref.~\cite{Tto0}).
This drastically reduces the rate of phase--slip: the $\phi$ variable
remains near the extrema of $\hat E$, and changing it with
$\hat f$ fixed (because $\tau_E$ is large) entails overcoming a
barrier of order $E_{\rm neq}$.  This surprising fact --- that noise
enhances conductance near $I=0$ --- is demonstrated numerically below.

\begin{figure}[t]
\epsfxsize=\hsize 
\epsffile{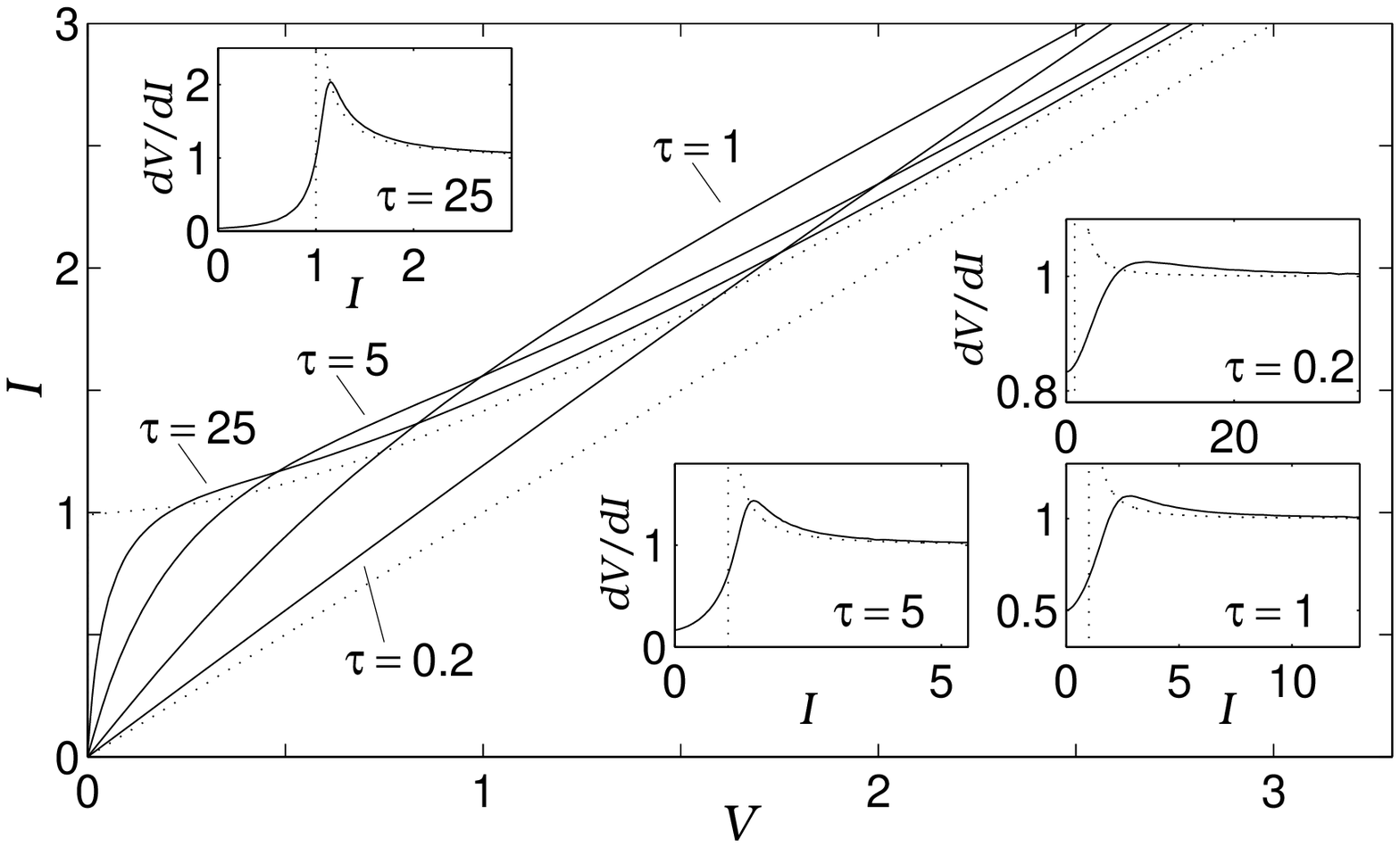}
\refstepcounter{figure}
\label{fig_IV}
FIG. \ref{fig_IV}:
The $I$---$V$ curves and differential resistances (insets) for a dc
current bias in the NEQ model, for four values of the relaxation time
$\tau_E$; compared to the RSJ model with $I_c = I_{\rm neq}$, and to 
$I = G_N V$ (dotted lines). 
\end{figure}

\begin{figure}[t]
\epsfxsize=\hsize 
\epsffile{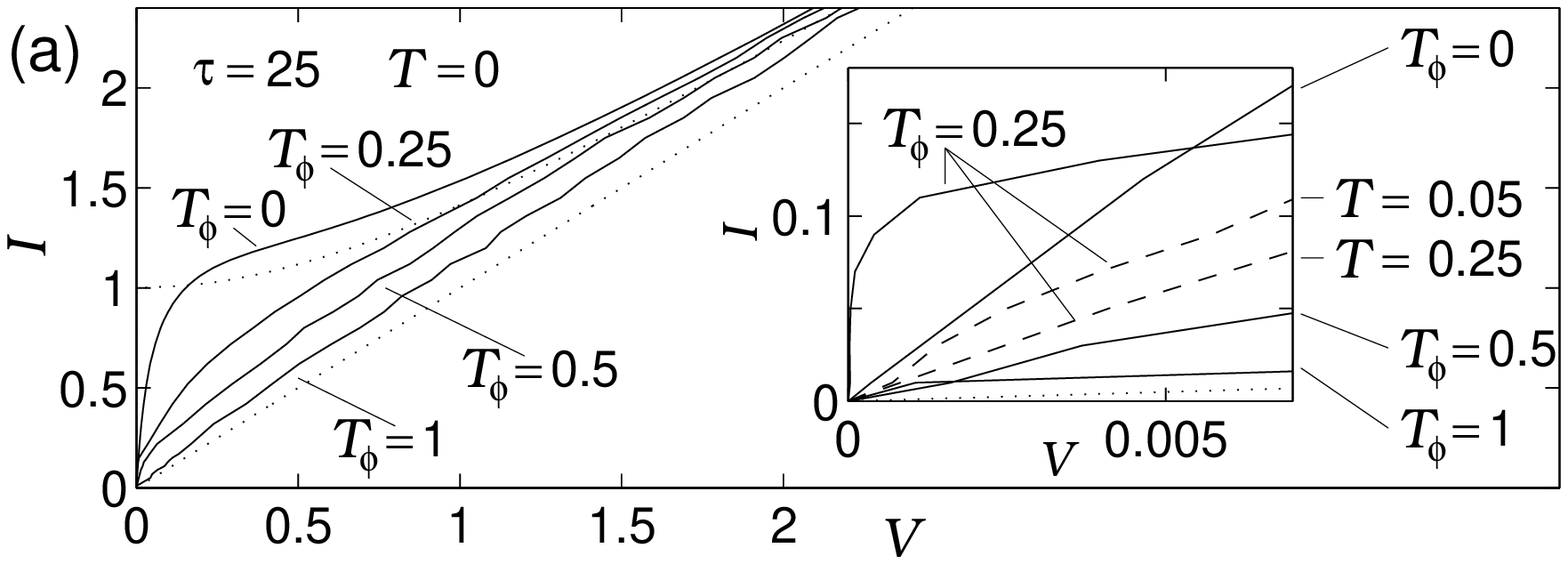}     
\epsfxsize=\hsize 
\epsffile{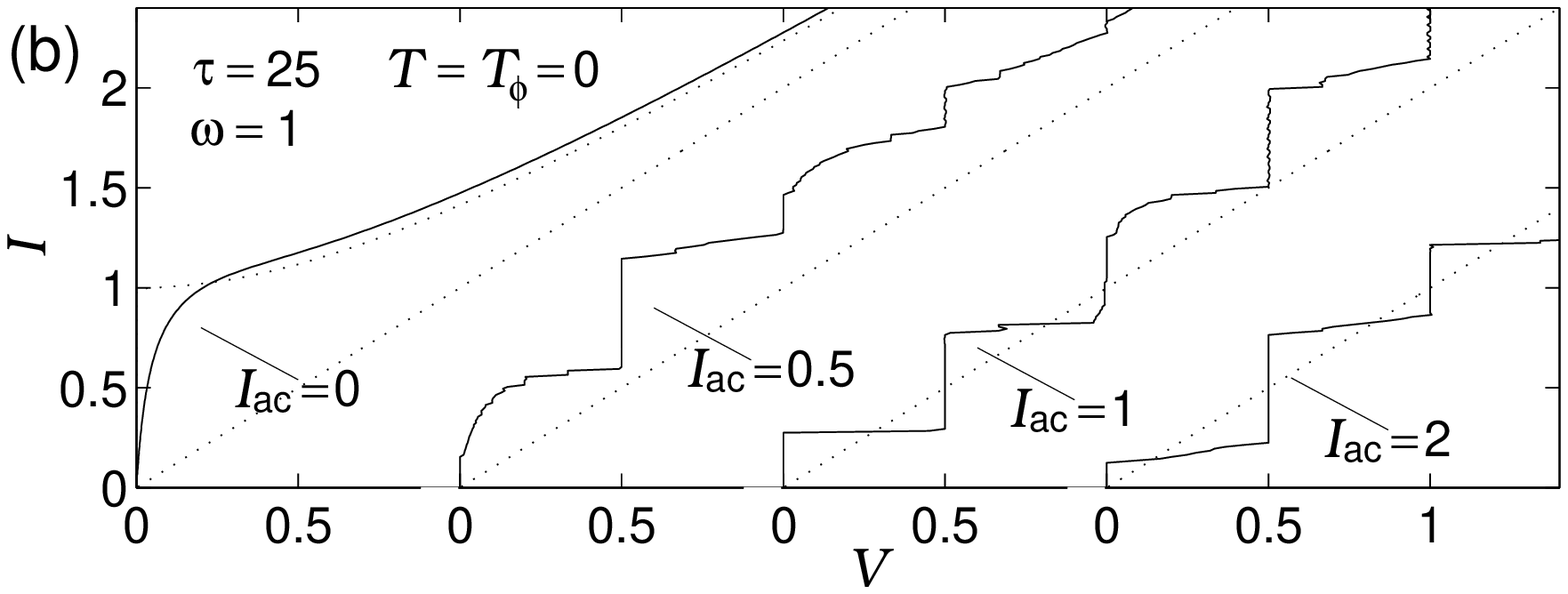}      
\vspace*{2mm}
\refstepcounter{figure}
\label{fig_shap}
FIG. \ref{fig_shap}:
(a) The effect of external noise (indicated by $T_\phi$) on the dc
$I$---$V$ curves with $\tau=25$ (the waviness is
due to the finite numerical integration time).
Inset: on a much expanded scale the $T_\phi = 0.25$ curve is seen to
cross the one for $T_\phi = 0$, and appears to have a ``critical
current'' of $\sim 0.1 I_{\rm neq}$.
Thermal noise counteracts this ($T > 0$; dashed lines).
(b) An ac current bias (with amplitudes $I_{\rm ac}$, frequency 
$\omega = 2eV_{\rm neq}/\hbar$, and no noise) produces Shapiro steps.
The step at $V=0.5$ is due to supercurrents at twice the Josephson
frequency.
\end{figure}

The dc $I$---$V$ curves of this model, obtained with 
$I(t) = {\rm const.}$ and no noise ($I_F = J_F =0$), are displayed in
Fig.~\ref{fig_IV},
using units of $I_{\rm neq}$, $V_{\rm neq}=G_N I_{\rm neq}$ 
and $\tau_{\rm neq} = \hbar/2eV_{\rm neq}$ for current, voltage and
time.  The results for large $\tau_E$ lie remarkably close
to the $I$---$V$ curves of the RSJ model \cite{fzbc}, except for the
finite zero--bias conductance, equal to $1+\tau$, or
$G_N + \tau_E (2e/\hbar) I_{\rm neq}$.  Results with external noise
(and a large $\tau_E$) are shown in Fig.~\ref{fig_shap}(a).
Such numerical calculations indicate that for finite $\tau_E$ the
scaled conductance grows at most as $\tau^2$, and not exponentially as
in the $\tau_E \to \infty$ case.

An ac drive, $I_{\rm ac} \cos(\omega t)$, produces Shapiro steps in
the $I$---$V$ curves --- phase--locking of the ac Josephson
oscillations to the external frequency $\omega$, see
Fig.~\ref{fig_shap}(b).  As in the RSJ model
\cite{Likharev}, closely--related results obtain for the
simpler, voltage--biased case \cite{unpub}.  For weak ac driving, the
sub--harmonic step at $\half \omega$ dominates.  As
observed in the experiments of Ref.~\cite{Allen}, this step is the
largest for $T \stackrel{>}{\sim} E_C$, but does not
appear to be parametrically larger than the others
(a mechanism with true frequency doubling would behave differently).  
The experimental steps were much 
smaller in size, which could indicate fast equilibration --- the
steps disappear quadratically for small $\tau_E$, because to leading
order \Eq{lowV} gives $I = G(\phi) V$, which relates the time
integrals of the current and voltage by a constant ratio, 
$\int G(\phi) \, d\phi/2\pi$, and precludes any steps
\cite{comment}.  A detailed modeling of the
experiment requires additional measurements, which are in progress.

In summary, we have discussed NEQ effects very similar to both the dc
(Fig.~\ref{fig_IV}) and the ac (Fig.~\ref{fig_ac}) Josephson effects,
which decay with temperature only as $I_{\rm neq} \propto E_C/T$.  
The NEQ occupations can be produced by microwave irradiation,
a noisy external circuit, or simply an applied voltage.  The
new effects have the signature of being dominated by the second
harmonic of the Josephson frequency, and may have already been
observed.

I would like to thank S.J. Allen, J.G.E. Harris, H. Kroemer,
K. Lehnert, B. Spivak, M. Tinkham, and F. Zhou for fruitful discussions. 
This work was supported by NSF grants No.\ PHY94-07194 and No.\
DMR96-30452.


\begin{references}
\bibitem{imbal} Situations of ``branch imbalance'' \cite{Tinkham}, for
which the occupations of the $E_n$ and the $-E_n$ levels are not
simply related, do not arise in the present work [see \Eq{relax}].
\bibitem{check}  To quantify the errors we have used the full set 
of $E_n(\phi)$ curves, and found deviations of less than 15\%,  for
85\% of the harmonic amplitudes shown in Fig.~\ref{fig_ac}.
\bibitem{Tto0} For $T \s= 0$, some noise in $d\hat f /dt$ is still induced
by $T_\phi$.
\bibitem{fzbc} In contrast to the RSJ model with noise, (which also
has $dV/dI \s> 0$ near $I \s= 0$ \cite{Likharev}) the different curves
of Fig.~\ref{fig_IV} cross each other --- as $\tau_E \to 0$ 
the maximum in $dV/dI$ shifts higher and higher, to 
$I \simeq G_N \hbar/e \tau_E \gg I_{\rm neq}$.
\end{references}
\end{document}